\documentclass[twocolumn,showpacs,preprintnumbers,amsmath,amssymb,floatfix,prc,
superscriptaddress]{revtex4}
\usepackage{graphicx}
\usepackage{dcolumn}
\usepackage{bm}
\def\la{\langle}
\def\ra{\rangle}
\def\beq{\begin{equation}}
\def\eeq{\end{equation}}
\def\be{\begin{eqnarray}}
\def\ee{\end{eqnarray}}
\def\hs{\hat{s}}
\def\htm{\hat{t}}
\def\hu{\hat{u}}

\newcommand{\f}[2]{\frac{#1}{#2}}
\newcommand{\dd}  { {\textrm d}}
\newcommand{\ba}{\begin{eqnarray*}}
\newcommand{\ea}{\end{eqnarray*}}

\usepackage{color}

\newcommand{\lsim}{
 \mathrel{\setbox0=\hbox{$<$}\raise0.6ex\copy0\kern-\wd0
 \lower0.65ex\hbox{$\sim$}}}

\newcommand{\gsim}{
 \mathrel{\setbox0=\hbox{$>$}\raise0.6ex\copy0\kern-\wd0
 \lower0.65ex\hbox{$\sim$}}}

\begin{document}
\title{Relativistic light-on-heavy nuclear collisions and the implied rapidity asymmetry}
\author{Adeola Adeluyi}
\affiliation{Center for Nuclear Research, Department of Physics \\
Kent State University, Kent, OH 44242, USA}
\author{Gergely G.~Barnaf\"oldi}
\affiliation{Center for Nuclear Research, Department of Physics \\
Kent State University, Kent, OH 44242, USA}
\affiliation{MTA KFKI RMKI Research Institute for Particle and Nuclear Physics\\
P.O. Box 49, Budapest 1525, Hungary}
\author{George Fai}
\affiliation{Center for Nuclear Research, Department of Physics \\
Kent State University, Kent, OH 44242, USA}
\author{P\'eter L\'evai}
\affiliation{MTA KFKI RMKI Research Institute for Particle and Nuclear Physics\\
P.O. Box 49, Budapest 1525, Hungary}
\date{\today}
\begin{abstract}
We calculate pseudorapidity ($\eta$) asymmetry in $pA$ and $dA$ collisions in 
a pQCD-improved parton model. With the calculations tuned to describe existing 
spectra from $pp$ collisions and asymmetric systems at midrapidity and large 
rapidities at FNAL and RHIC energies, we investigate the roles of nuclear 
shadowing and multiple scattering on the observed asymmetry. Using this framework, 
we make predictions for pseudorapidity asymmetries at high $p_T$ and large $\eta$ 
in a wide range of energies up to LHC.
\end{abstract}
\maketitle
\vspace{1cm}
%
%
\section{Introduction}
\label{intro}

Collisions of asymmetric nuclear systems, like proton-nucleus or deuteron-nucleus 
collisions, attract significant experimental and theoretical attention at present.  
At the Relativistic Heavy Ion Collider (RHIC), Run 8 included deuteron-gold
collisions at 200 $A$GeV with a luminosity increase of about an order of magnitude
compared to the deuteron-gold run of 2003. One of the physics goals was to provide 
a high-statistics ``cold'' nuclear matter data set to establish a definitive baseline 
for ``hot'' nuclear matter, created in gold-gold collisions. The benchmark role of the
``the deuteron-gold control experiment'' for e.g. energy-loss studies has often been 
emphasized~\cite{Gyulassy:2004jj,Hemmick:2004jc}.
Proton-nucleus and deuteron-nucleus reactions were 
also used to study the Cronin-effect~\cite{Cron75,Antr79}.  

Light-on-heavy nucleus-nucleus collisions offer unique information about the 
underlying dynamics, not available in symmetric proton-proton or 
nucleus-nucleus collision systems. The light-on-heavy asymmetry
manifests itself in an asymmetric (pesudo-)rapidity distribution of 
charged particles with respect to
zero rapidity (or pseudorapidity) as measured by BRAHMS~\cite{Arsene:2004cn} and 
PHOBOS~\cite{Back:2004mr}. The asymmetry of the yields can be quantified by 
introducing the ratio of pseudorapidity densities at a given negative 
pseudorapidity relative to that at
the positive pseudorapidity of the same magnitude. This backward/forward ratio
(forward being the original direction of motion of the light partner) is referred to
as pseudorapidity asymmetry. The STAR Collaboration published pseudorapidity 
asymmetries in 200 $A$GeV $dAu$ collisions for several identified hadron species 
and total charged hadrons in the pseudorapidity intervals $|\eta| \le 0.5$ and 
$0.5 \le |\eta| \le 1.0$~\cite{Abelev:2006pp}. Asymmetries with the
backward/forward ratio above unity for transverse momenta up to $\approx 5$ GeV/c
are observed for charged pion, proton+anti-proton, and total charged hadron production 
in both rapidity regions. Because of the importance of asymmetric collisions, we 
anticipate that proton-lead or deuteron-lead data will be 
collected at the Large Hadron Collider (LHC). 

The theoretical relevance of asymmetric collision systems (in particular colliding
the lightest nuclei like protons or deuterons on a heavy partner) in 
testing parameterizations for nuclear shadowing and models for initial multiple 
scattering was recognized prior to the availability of 200 $A$GeV RHIC deuteron-gold 
data~\cite{Vitev:2003xu,Wang:2003vy,Barnafoldi:2004kh}. One can take advantage
of the fact that
positive (forward) rapidities correspond to large parton momentum fractions from
the light partner and small momentum fractions from the heavy nucleus
(and vice versa for negative rapidities), and that there are more collisions 
suffered traversing the heavy partner. Wang made predictions for pseudorapidity 
asymmetries on this basis~\cite{Wang:2003vy}. A subsequent calculation focused on the 
lowest transverse momenta where pQCD may be applicable~\cite{Barnafoldi:2005rb}.
More recently, two versions of a pQCD-based model with nuclear modifications were 
used to address deuteron-gold collisions and results were compared to available
data~\cite{Adeluyi:2008qk,Barnafoldi:2008rb}. The related subject of
forward-backward rapidity correlations in asymmetric systems is treated in
~\cite{Brogueira:2007ub} using a derivative of string percolation models, and 
in \cite{Armesto:2006bv} under the framework of color glass condensate.  

In the present study we investigate the roles of nuclear shadowing and multiple 
scattering in the generation of rapidity asymmetry 
in the backward/forward yields
at intermediate and high transverse-momenta. Here we use the HIJING
shadowing parameterization~\cite{Li:2001xa} and the recently released 
Eskola--Paukkunen--Salgado (EPS08) nuclear parton distribution functions
(nPDFs)~\cite{Eskola:2008ca}. While the former has been applied widely, 
the latter was not available at the time of the
earlier studies mentioned above. We calculate the pseudorapidity asymmetry in three 
representative asymmetric light-on-heavy systems: $pBe$ at $30.7$ GeV (Fermilab), 
$dAu$ at $200$ $A$GeV (RHIC), and $dPb$ at $8.8$ $A$TeV (LHC). 
We concentrate on neutral pion production and compare results to experimental data 
from the E706 experiment~\cite{Apanasevich:2002wt},
PHENIX~\cite{PHENIXdAu}, and STAR~\cite{Abelev:2006pp}.

The paper is organized as follows: in Sec.~\ref{col} we review the basic 
formalism of the pQCD-improved parton model with intrinsic transverse momentum and
multiple scattering as applied to proton-nucleus ($pA$) and deuteron-nucleus ($dA$) 
collisions. This section also includes the definition of the pseudorapidity asymmetry 
and a discussion of the roles of nuclear shadowing and multiple scattering in asymmetric 
collisions. We present the results of our calculation in Sec.~\ref{res} and conclude 
in Sec.~\ref{concl}.

\section{Calculational framework}
\label{col}

\subsection{Description of the model}
\label{partmd}
The invariant cross section for the production of 
final hadron $h$ from the collision of nucleus $A$ and nucleus $B$ 
($A+B \!\to\! h+X$), can be written, in collinearly factorized pQCD, as
\begin{eqnarray}\label{eq:pdA}
E_h\f{\dd ^3 \sigma_{AB}^{h}}{\dd^3 p} =
\sum_{\!\!abcd}\!\int\!\! \dd^2b\ \dd^2r\ t_A(b) t_B(|\vec{b}-\vec{r}|)\dd{x_a} \dd{x_b}
\nonumber \\ 
\dd{z_c} \,
f_{\!a/A}(x_a,Q^2)\ f_{\!b/B}(x_b,Q^2)\
\nonumber \\ 
\f{\dd\sigma(ab\!\to\!cd)}{\dd\htm}\,
\frac{D_{h/c}({z_c},\!Q_f^2)}{\pi z_c^2}
\hs \, \delta(\hs\!+\!\htm\!+\!\hu)   \,\,  ,
\end{eqnarray}
where $x_a$ and $x_b$ are parton momentum fractions in $A$ and $B$, respectively, 
and $z_c$ is the fraction of the parton momentum carried by the final-state 
hadron~$h$. 
The factorization and fragmentation scales are $Q$ and $Q_f$, respectively. 
As usual, $\hs$, $\htm$, and $\hu$ refer to the partonic Mandelstam variables, 
the massless parton approximation is used, and
\begin{equation} \label{eq:tf}
t_{A}(\!{\vec s}) = \int\!\! \dd z \rho_{A}(\!{\vec s},z)
\end{equation}
is the Glauber thickness function of nucleus $A$, with the nuclear density distribution,
$\rho_{A}({\vec s},z)$ subject to the normalization condition
\begin{equation}
\int\!\! \dd^2s \ \dd z \rho_{A}({\vec s},z) = A \,\, .
\end{equation}
The quantity $\dd\sigma(ab\!\to\!cd)/\dd\htm$ in eq.~(\ref{eq:pdA}) represents the 
perturbatively calculable partonic cross section, and $D_{h/c}(z_c,\!Q_f^2)$ stands 
for the fragmentation function of parton $c$ to produce hadron $h$, evaluated at momentum 
fraction $z_c$ and fragmentation scale $Q_f$. 

The collinear parton distribution functions (PDFs) can be generalized to include a transverse
momentum degree of freedom, ${\vec k}_T$, as required by the uncertainty principle. This
can be formally implemented in terms of unintegrated PDFs~\cite{Collins:2007ph,Czech:2005vy}. 
To avoid some of the complications associated with using unintegrated
PDFs, it is expedient to parameterize phenomenologically the ${\vec
  k}_T$ dependence of the parton distributions. In this light   
a phenomenological model for proton-proton ($pp$), proton-nucleus ($pA$), 
and nucleus-nucleus ($AB$) collisions incorporating parton transverse 
momentum in the collinear pQCD formalism was developed in
Ref.~\cite{Zhang:2001ce}. In order to make the present study
reasonably self-contained, we include details relevant to the
present treatment.

In this model, the invariant cross section can be written as
\begin{eqnarray}\label{eq:pdA_kt}
E_h\f{\dd ^3 \sigma_{AB}^{h}}{\dd^3 p} =
\sum_{\!\!abcd}\!\int\!\! \dd^2b\ \dd^2r\ t_A(b) t_B(|\vec{b}-\vec{r}|)\dd{x_a} \dd{x_b}
\nonumber \\ 
 \dd{\vec k}_{Ta} \, \dd{\vec k}_{Tb} \, \dd{z_c} \,
f_{\!a/A}(x_a,\!{\vec k}_{Ta},Q^2)\ f_{\!b/B}(x_b,\!{\vec k}_{Tb},Q^2)\
\nonumber \\ 
\f{\dd\sigma(ab\!\to\!cd)}{\dd\htm}\,
\frac{D_{h/c}({z_c},\!Q_f^2)}{\pi z_c^2}
\hs \, \delta(\hs\!+\!\htm\!+\!\hu)   \,\,  ,
\end{eqnarray}
where the $k_T$-broadened parton distribution in the nucleon is
written, in a simple product approximation, as
\begin{equation}
f_{a/N}(x,{\vec k}_{T},Q^2) \longrightarrow  \ g({\vec k}_T) \cdot  f_{a/N}(x,Q^2)  \,\, ,
\end{equation} 
with  $f_{a/N}(x,Q^2)$ denoting the standard collinear PDF in the nucleon.
The transverse momentum distribution is taken to be
a Gaussian,
\begin{equation}\label{eq:kT}
g({\vec k}_T)  =  \frac{\exp(-k_T^2/\langle k_T^2 \rangle_{pp})} {\pi \langle k_T^2 \rangle_{pp}} \,\, ,
\end{equation}
where $\langle k_T^2 \rangle_{pp}$ is the two-dimensional width of the transverse-momentum
distribution in the proton.
Based on the then-available pion and unidentified hadron production data, an estimate 
for the model width of the transverse-momentum distribution of partons in the proton
($\langle k_T^2 \rangle_{pp}$) 
was presented in Ref.~\cite{Zhang:2001ce}.
A recent summary of the energy dependence of the model parameter $\langle k_T^2 \rangle$, 
related to the average transverse momentum of the created pair by
\begin{equation}\label{eq:pT_kT}
\langle k_T^2 \rangle_{pp} = \frac{\langle p_T \rangle^{2}_{pair}}{\pi}  \,\, ,
\end{equation}
can be found in Ref.~\cite{Barnafoldi:2007uw}. The value of $\langle k_T^2 \rangle_{pp}$
increases logarithmically with $\sqrt{s}$, and the data are well described by the function
\begin{equation}\label{eq:pT_rs}
\langle p_T \rangle_{pair} = (1.74 \pm 0.12) \centerdot \log_{10}(\sqrt s) +
(1.23 \pm 0.2)  \,\, .
\end{equation}
Using eq.~(\ref{eq:pT_rs}) one can estimate 
$\langle k_T^2 \rangle$ at required cms energies.
It should be noted that eq.~(\ref{eq:pT_rs}) is a fit to the data at
lower energies. Since no data exists at LHC energies, we
extrapolate $\langle k_T^2 \rangle$. We observe that: 
(a) Since the dependence is logarithmic, the energy step 
we make from RHIC to LHC is comparable to the one from SPS to RHIC.
We thus do not a priori expect a radical change in the trend.
(b) The $k_T$ effects have been shown to be appreciable at low $p_T$, and since 
our predictions at LHC cover a wide $p_T$ range up to hundreds of GeV/c in
$p_T$, potential uncertainties arising from the extrapolation of Eq. 8 will 
affect a small fraction of this range at the low $p_T$ end.
(c) The present study, at LHC energies, is exploratory. Results of pp
collisions at the LHC will be helpful in determining the magnitude of
the presently uncertain $\langle k_T^2 \rangle_{pp}$.

It is easy to appreciate the physical necessity of the presence of a 
transverse-momentum degree of freedom in proton-proton and, therefore, nuclear 
collisions. However, the handling of the effect of the nuclear environment 
on transverse momenta is one of the specific features of the given description. 
Most shadowing parameterizations include at least some of the effects
of multiple scattering in the nuclear medium, while the HIJING parameterization
(as we discuss further in Sec.~\ref{shad_multiscatt}) needs to be augmented
with modeling nuclear multiscattering. For this purpose, in $pA$ collisions
we use a broadening of the width of the transverse momentum distribution 
(\ref{eq:kT}) according to 
\begin{equation}\label{eq:broad}
\langle k_T^2\rangle_{pA} = \langle k_T^2\rangle_{pp} + C\ h_{pA}(b) \,\, ,
\end{equation}
where $\langle k_T^2\rangle_{pp} $ is the width already present in proton-proton 
collisions, $h_{pA}(b)$ is the number of effective nucleon-nucleon ($NN$) collisions as a 
function of nucleon impact parameter $b$, and $C$ is the average increase in width 
per $NN$ collision. In nucleus-nucleus ($AB$) collisions, the $p_T$ distributions of
both nuclei are subject to the broadening represented by eq.~(\ref{eq:broad}). 

The function $h_{pA}(b)$ can be written in terms of the number of collisions suffered 
by the incoming proton in the target nucleus, $\nu_{A}(b) =
\sigma_{NN}t_{A}(b)$, where $\sigma_{NN}$ is the inelastic nucleon-nucleon 
cross sections. It was found in 
Ref.~\cite{Zhang:2001ce} that only a limited number of collisions is effective in broadening
the transverse momentum distribution. This model scenario was referred to as
``saturation'', and an optimal description was found with
the effective number of nucleon-nucleon collisions maximized at $4$, 
and the average width increase per $NN$ collision, $C$, set to $0.35$  GeV$^2/c^2$.
We do not change the values of these parameters in the present study. The resulting
transverse momentum broadening may appear too large relative to what is observed 
in Drell-Yan data at FNAL. However, we focus on meson production in the present
application, where the PHENIX experiment extracts intrinsic transverse momenta
in $pp$ collisions at RHIC energies that are similar in magnitude to the ones
discussed here\cite{Adler:2006sc}.
Further details about this aspect of the model
can be found in Ref.~\cite{Zhang:2001ce}.

The intrinsic transverse momentum $k_T$ is treated 
phenomenologically in Ref.~\cite{Zhang:2001ce}  and in the 
present study. While next-to-leading-order (NLO) calculations provide
a more accurate description of the parton-level cross section, they continue
to rely on the factorization theorem and represent the non-perturbative
information in terms of PDFs and fragmentation functions. Since these
functions are fitted to the same data as in LO, the expected change is a shift
in responsibilty between the perturbative and non-perturbative sectors in
describing the data used to define the non-perturbative ingredients. In 
aplications to a larger set of data this will not eliminate the need for the 
phenomenological use of a transverse momentum distribution, in particular
considering the fact that even NLO may be far from a full perturbative
expansion. In addition, above we argue for inclusion of a transverse momentum 
degree of freedom on fundamental physical grounds as basic as the uncertatinty
principle. The simple Gaussian representation of this physics provides a 
phenomenologically useful additional parameter (the width). This holds true
both at the LO and NLO levels. Going to NLO may change the range of transverse
momenta where intrinsic $k_T$ is important (see e.g. 
Refs.~\cite{MRTS:1995,CTEQ2:1995,Vogelsang:2007}).
No attempt is made in the present work to discuss the $x$,  
flavor (quarks and gluons), and rapidity dependence of $k_T$.  We have 
limited ourselves here to a simple effective description. This seems adequate for now, 
since $k_T$ effects are appreciable only at relatively low $p_T$, and a 
major focus of this work is asymmetry at high $p_T$ (except at very forward 
rapidities where $p_T$-s are low due to phase space constraints). The intrinsic 
transverse momentum, $\langle k_T^2\rangle_{pp}$, enhances hadron production 
yields at low $p_T$ in both negative and positive pseudorapidity regions, thus tending 
to cancel out in the asymmetry ratio. Multiscattering, on the other hand, does impact 
appreciably on the asymmetry even at low $p_T$.
Overall, calculations within this framework 
have proven their value in the interpretation of hadron-production 
data~\cite{Zhang:2001ce,Levai:2003at}. (For a comparison of the leading-order 
$k_T$-factorized approach and next-to-leading order collinear approach see 
Ref.~\cite{Szczurek:2007bt}.)

The collinear nPDFs $f_{a/A}(x,Q^2)$ are 
expressible as convolutions of nucleonic parton distribution functions (PDFs)
$f_{a/N}(x,Q^2)$ and a shadowing function ${\cal S}_{a/A}(x,Q^2)$ which encodes the 
nuclear modifications of parton distributions.  We use the MRST2001
PDFs~\cite{Martin:2001es} for the nucleon 
parton distributions, and for the shadowing function we employ both 
the EPS08 shadowing routine~\cite{Eskola:2008ca} and HIJING~\cite{Li:2001xa}.
(Other nPDFs, like FGS~\cite{Frankfurt:2003zd}, HKN~\cite{Shad_HKN}, 
and the earlier EKS~\cite{Eskola:1998df}, are used elsewhere
to calculate pseudorapidity asymmetries~\cite{Adeluyi:2008qk}.) 
For the final hadron fragmentation we utilize the fragmentation functions
in the AKK set~\cite{Albino:2005me}. The factorization scale is tied to the 
parton transverse momentum via $Q = (2/3) p_T/z_c$, while the
fragmentation scale varies with the transverse momentum of the outgoing hadron,
according to $Q_f = (2/3) p_T$, following the best-fit results obtained in
Ref.~\cite{Levai:2006yd}. This scale fixing is not modified from earlier applications,
and is used consistently throughout the present calculations.
To protect against divergences in the partonic cross sections, a cutoff 
regulator mass is necessary. We have tested the sensitivity of our
calculations at different cms energies by varying the regulator mass 
between $0.5-2.0$ GeV. For $p_T>1.5$ GeV/c our results show little
sensitivity to variation of regulator mass. We present results for 
$p_T>2.25$ GeV/c here, and thus regulator mass effects are expected 
to be minimal.
We obtain the density distribution of the deuteron from the Hulthen 
wave function~\cite{Hulthen1957} (as in Ref.~\cite{Kharzeev:2002ei}), while a 
Woods-Saxon density distribution is used for gold and lead with parameters from
Ref.~\cite{DeJager:1974dg}.

%
\subsection{Forward and backward nuclear modifications}
\label{rapasym}

The nuclear modification factor is designed to compare, as a ratio, spectra 
of particles produced in nuclear collisions to a hypothetical scenario 
in which the nuclear collision is assumed to be a superposition of the 
appropriate number of nucleon-nucleon collisions. The ratio can be 
defined as a function of $p_T$
for any produced hadron species $h$ at any pseudorapidity $\eta$:
\begin{equation}
R^h_{AB}(p_T, \eta) = \frac{1}{\langle N_{bin}\rangle} \cdot
\frac{E_h \dd^3\sigma_{AB}^{h}/\dd^3 p |_{\eta}}
{E_h \dd^3\sigma_{pp}^{h}/\dd^3 p |_{\eta}}   \,\, ,
\label{rdau}
\end{equation}
where $\la N_{bin} \ra$ is the average number of binary collisions in 
the various impact-parameter bins.
Nuclear effects manifest themselves in $R^h_{AB}(p_T, \eta)$ values greater or 
smaller than unity, representing enhancement or suppression, respectively, 
relative to the $NN$ reference.

In asymmetric collisions, hadron production at forward rapidities
may be different from what is obtained at backward 
rapidities. It is thus of interest to study ratios of particle
yields between a given pseudorapidity value and its negative in these
collisions. The pseudorapidity asymmetry $Y_{Asym}(p_T)$ is defined 
for a hadron species $h$ as 
\begin{equation}
Y^h_{Asym}(p_T) = \left. E_h\f{\dd ^3 \sigma_{AB}^{h}}{\dd^3 p} \right|_{-\eta} 
 \left/ 
\left. E_h\f{\dd ^3 \sigma_{AB}^{h}}{\dd^3 p} \right|_{\eta} \right. \,\, .
\label{yasym}
\end{equation}

Let us consider the (double) ratio of the  
forward and backward nuclear modification factors in $dAu$
collisions for species $h$:
\begin{eqnarray}
R^h_{\eta}(p_T) = \frac{R^h_{dAu}(p_T,-\eta)}{R^h_{dAu}(p_T,\eta)}= \nonumber \\  
\frac{E_h \dd^3\sigma_{dAu}^{h}/\dd^3 p |_{-\eta}}
{E_h \dd^3\sigma_{pp}^{h}/\dd^3 p |_{-\eta}} \left/  
\frac{E_h \dd^3\sigma_{dAu}^{h}/\dd^3 p |_{\eta}}
{E_h \dd^3\sigma_{pp}^{h}/\dd^3 p |_{\eta}} \right. \,\, . 
\label{y-r:eq}
\end{eqnarray}
As discussed in Ref.~\cite{Barnafoldi:2008rb}, since the $pp$ 
rapidity distribution is symmetric around $y=0$,
if the same backward and forward (pseudo)rapidity ranges are taken
in both directions (i.e. $ |\eta_{min}| \leq |\eta| \leq |\eta_{max}| $),  
then the $pp$ yields cancel in eq.~(\ref{y-r:eq}) and one obtains
that the ratio defined in (\ref{y-r:eq}) is identical to 
the pseudorapidity asymmetry (\ref{yasym}):  
\begin{equation}
Y^h_{Asym}(p_T)= R^h_{\eta}(p_T)=\frac{R^h_{dAu}(p_T,-\eta)}{R^h_{dAu}(p_T,\eta)} \,\, .
\label{y-r2:eq}
\end{equation}

Eq. (\ref{y-r2:eq}) is useful from the experimantal point of view to handle the 
systematic errors of data and extract the proper pseudorapidity asymmetry. It also 
provides a connection between measured rapidity asymmetry and the nuclear 
modification factors.

\subsection{Nuclear shadowing and multiple scattering}
\label{shad_multiscatt}

We expect that particle production in $pA$ ($dA$) collisions will have 
different yields in the forward and backward directions. This is because the
respective partons have different momentum fractions (shadowing differences)
and because the forward-going parton has to traverse a large amount of matter.
We found in our earlier studies that the HIJING shadowing parameterization
is particularly useful for the study of multiple scattering, as it requires
an explicit treatment of $k_T$ broadening (see eq.~(\ref{eq:broad})) for a 
successful description of data~\cite{Zhang:2001ce}. In this Section we therefore 
study the interplay of shadowing and multiple scattering in more detail 
using the HIJING parameterization. 

\begin{figure}[!htb]
\begin{center} 
\includegraphics[width=8.5cm, height=8.5cm, angle=0]{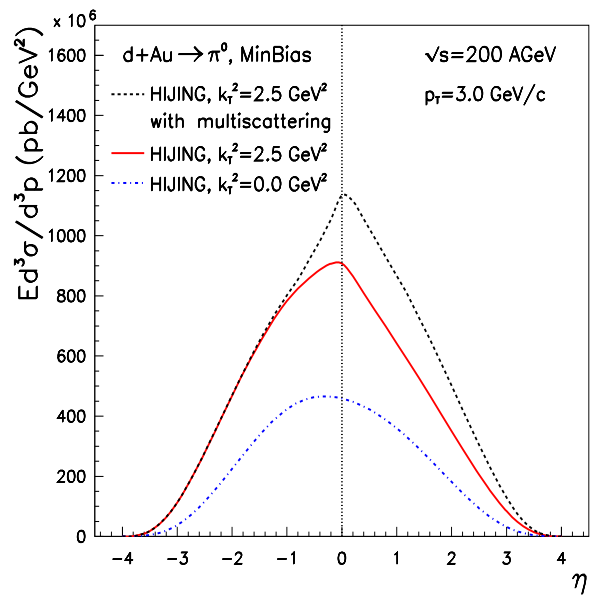}
\end{center}
\caption[...]{(Color Online) Illustration of calculated yields at $p_T=3.0$ GeV/c in
minimum bias neutral pion production from $dAu$ collisions at 200 $A$GeV.  
The solid line represents HIJING shadowing with intrinsic $k_T$ in the
proton, the dashed line is HIJING with intrinsic $k_T$ and multiple 
scattering, while the dot-dashed curve is obtained by turning off
any transverse momentum.}
\label{fig:multiscatt}
\end{figure}

In Fig.~\ref{fig:multiscatt} we display the 
neutral pion yield as a function of pseudo-rapidity in $dAu$ collisions 
at a fixed $p_T$, namely at $ 3.0$ GeV/c. This illustrates how the 
nuclear effects we examine modify the asymmetry of the yields. 
The figure shows the calculated distribution using the HIJING shadowing
parameterization without any intrinsic $k_T$ (dot-dashed curve), 
using HIJING shadowing and including the intrinsic $k_T$ in the nucleon (solid), 
and HIJING shadowing with intrinsic $k_T$ and multiple scattering (dashed).
The parameters of eq.~(\ref{eq:broad}) are unchanged from our previous 
studies at midrapidity~\cite{Levai:2003at,Levai:2006yd}, and 
we have chosen a transverse momentum value comparable to $\la k_T \ra_{pp}$,
where the various effects are clearly displayed.  

We checked that when shadowing and all nuclear effects are turned off, 
the distribution is symmetric around midrapidity, as expected. 
With shadowing only, we find an asymmetry in the distribution: the yield 
at a fixed negative pseudorapidty, say $\eta=-2$ ($Au$ side or ``backward'') 
is higher than at the corresponding positive pseudorapidity ($\eta=2$, $d$ side,  
or ``forward''). Thus the pseudorapidity asymmetry, $Y_{Asym}$, is greater 
than unity in this case. The inclusion of the intrinsic transverse momentum 
in the proton significantly increases the yields on both the $Au$ side and the
$d$ side at this transverse momentum, as it makes larger $p_T$-s accessible.
We tested that adding intrinsic transverse momentum in the proton in the 
absence of shadowing does not destroy the forward/backward symmetry. When
intrinsic $k_T$ in the proton is added to the calculation using shadowing,
but without the multiple scattering contribution, the sense of the asymmetry 
given by shadowing (backward/forward $\geq$ 1) is preserved. However, when 
multiple scattering is included, the yield shows a stronger increase in 
the forward direction. This is understandable in the present picture, since 
forward-going products originate in the partons of the deuteron, and have
to traverse a large amount of nuclear matter resulting in strong multiple 
scattering, while the backward products suffer no or little multiple 
scattering. This has the effect of reversing the asymmetry: the yield on the 
$d$-side is now greater than that on the $Au$-side. The calculated pseudorapidity 
asymmetry, $Y_{Asym}$, turns out to be less than unity in this case.
An alternative way to summarize the situation is to say that shadowing suppresses
the yield more on the $d$ side (forward) relative to a symmetric collision,
while the multiple scattering contribution is understandably large in the 
forward direction.
 
We have carried out similar studies at higher $p_T$ values. When the intrinsic
transverse momenta are small relative to the $p_T$ of the final hadron, $k_T$
effects become naturally smaller. Shadowing effects also become smaller as
the antishadowing region of the HIJING parameterization is approached. Thus,
at $p_T \gtrsim 15$ GeV/c the influence of multiple scattering and intrinsic
$k_T$ become negligible. These phenomena are most important at intermediate $p_T$ 
values, 2 GeV/c $\lesssim p_T \lesssim$ 8 GeV/c at RHIC. It is interesting to note
that a similar transverse-momentum region is sensitive to nuclear effects 
at lower (e.g. CERN SPS) energies, due to the $\sim \log(\sqrt{s})$ scaling of the 
Cronin peak~\cite{Barnafoldi:2007uw,e706,Zielinski}. 

%
\section{Results}
\label{res}

To judge the success of the model in reproducing spectra,  
here we first present calculated spectra for neutral pion production 
at midrapidity and non-zero rapidities. This will help select the model 
choices providing the best agreement with the experimental information. 
The selected model variants are then used to calculate pseudorapidity 
asymmetries.
 
\subsection{Midrapidity spectra}
\label{centrap}

\begin{figure}[!h]
\begin{center} 
\includegraphics[width=8.5cm, height=8.5cm, angle=270]{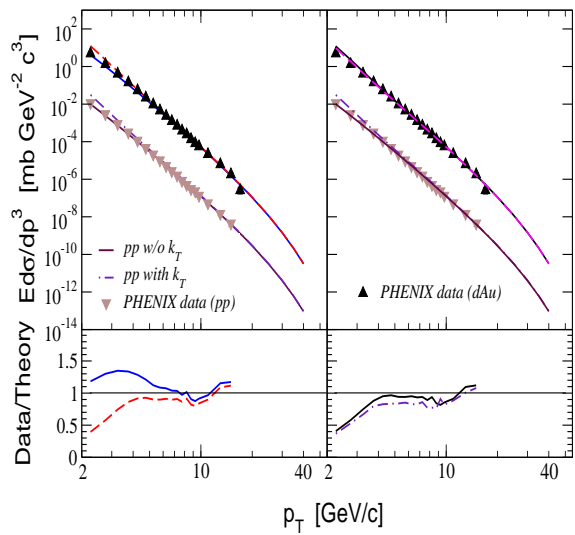}
\end{center}
\caption[...]{(Color Online) Top: spectra for $dAu$ $\pi^0$ production at 
$|\eta| < 0.35$. The left panel shows EPS08 with (dashed) and without (solid) 
intrinsic $k_T$. The right panel is HIJING plus $k_T$ with (solid) and without 
(dashed) multiple scattering. Filled triangles denote the PHENIX data~\cite{PHENIXdAu}. 
The $pp$ spectra from the same experiment and calculated with (dashed) and without 
(solid) intrinsic $k_T$ are also included. Bottom: corresponding data/theory
ratios for (left) $dAu$ with and without $k_T$  and (right) $pp$ with $k_T$
and HIJING plus $k_T$ with multiple scattering.}
\label{fig:dAuspect1}
\end{figure}

Figure~\ref{fig:dAuspect1} displays midrapidity RHIC $\pi^0$ spectra from 
$dAu$ collisions at $|\eta| < 0.35$, and for reference, the spectra 
from $pp$ collisions at the same energy ($\sqrt{s}= 200 A$GeV). The results are 
compared to data from the PHENIX collaboration~\cite{PHENIXdAu}. In the top left
panel we show calculated $dAu$ spectra using EPS08 nPDFs with and without intrinsic $k_T$ 
in the proton, but without any multiple scattering contribution. The top right depicts 
the $dAu$ spectra using HIJING plus $k_T$ with and without multiple scattering.
The bottom panels contain data/theory (i.e. data/model) ratios. 
As can be seen clearly in the bottom left $dAu$ data/theory panel, including 
intrinsic $k_T$ in the proton increases the calculated yield mostly at low $p_T$.
We consider the EPS08 description with intrinsic $k_T$ in the proton
satisfactory in the 4 GeV/c 
$\lesssim p_T \lesssim$ 10 GeV/c interval. Comparing the two bottom 
data/theory panels it is easy to see that the HIJING parameterization gives
a very similar accuracy for $dAu$ when multiple scattering is included. This is 
also very close to the data/theory for $pp$ collisions given by HIJING. 

\begin{figure}[!h]
\begin{center} 
\includegraphics[width=8.5cm, height=8.5cm, angle=270]{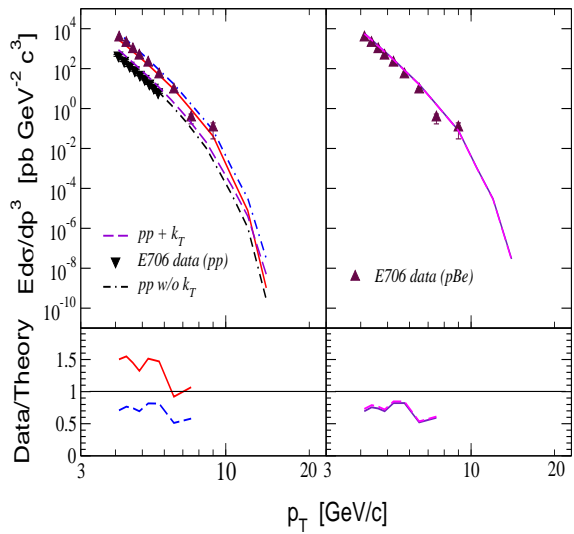}
\end{center}
\caption[...]{(Color Online) Spectra for $p+Be \rightarrow \pi{^0}+X$ at $|\eta| < 0.2$.
The left panel depicts EPS08 calculations with (dashed) and without (solid) intrinsic $k_T$. 
The right panel is HIJING plus $k_T$ with (solid) and without (dashed)
multiple scattering. Filled triangles denote the E706
data~\cite{Apanasevich:2002wt}. Data and calculations for proton-proton (pp)
collisions are at $530$ GeV/c.}
\label{fig:pBespect1}
\end{figure}

Fig.~\ref{fig:pBespect1} shows the spectra for neutral pion 
production from $pBe$ collisions at $|\eta| < 0.2$. The top left 
panel shows the spectra using EPS08 nPDFs with and without intrinsic
$k_T$. The top right depicts the spectra using HIJING plus $k_T$ 
with and without multiple scattering. The agreement with the E706 
data from Fermilab is quite good as can be seen from the lower panels
which display the data per theory ratio. The HIJING 
parameterization with and without multiple scattering gives almost 
identical data/theory ratios, i.e. multiple scattering has only
a small effect on spectra from $pBe$ 
collisions. This is reasonable in view of the fact that in 
the light $Be$ nucleus there are few 
scattering centers. The effect of intrinsic 
$k_T$ is to increase the calculated yield, leading to data/theory ratios
less than unity for all relevant $p_T$. 

We conclude from Fig.s~\ref{fig:dAuspect1} and \ref{fig:pBespect1} that
it is necessary to include the intrinsic transverse 
momentum of partons in the proton in our model to obtain a satisfactory
description of available data. The EPS08 and HIJING parameterizations differ
to the extent that while HIJING calls for the inclusion of the broadening of
the intrinsic transverse momentum distribution via multiple scattering, 
EPS08 appears to incorporate this physics in their nPDFs. We thus 
concentrate on three model variants (EPS08 with proton intrinsic $k_T$ and 
HIJING with intrinsic $k_T$ and with/without multiscattering) in the remainder
of this study.

%
\subsection{Spectra at non-zero rapidities}
\label{noncentrap}

\begin{figure}[!h]
\begin{center} 
\includegraphics[width=8.5cm, height=8.5cm, angle=270]{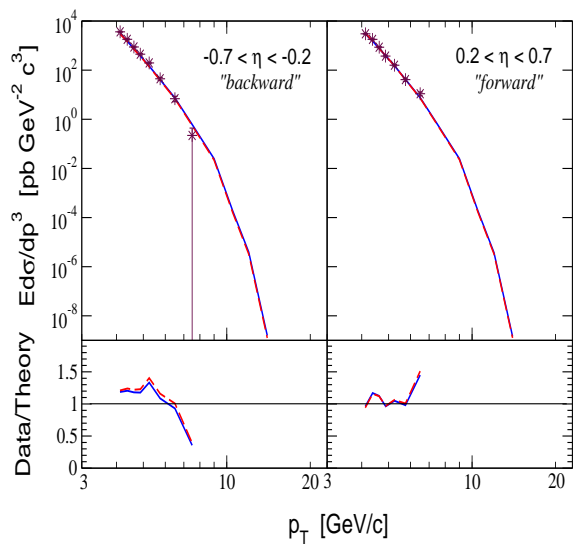}
\end{center}
\caption[...]{(Color Online) Spectra for $p+Be \rightarrow \pi{^0}+X$ at 
$-0.7 < \eta < -0.2$ (left panel) and $0.2 < \eta < 0.7$ (right panel). 
The solid lines represent the EPS08 nPDFs with intrinsic $k_T$, while 
the dashed line is obtained from HIJING 
plus $k_T$ with the inclusion of multiscattering. Stars denote 
the E706 data~\cite{Apanasevich:2002wt}.}
\label{fig:pBespect2}
\end{figure}

We now consider spectra at non-zero pseudorapidities. 
We display results from EPS08 with proton 
intrinsic $k_T$ and HIJING with intrinsic $k_T$ and multiscattering.
Fig.~\ref{fig:pBespect2} shows the spectra and corresponding data/theory ratio 
for $p+Be \rightarrow \pi{^0}+X$ at $-0.7 < \eta < -0.2$ (``backward'', left panel) 
and $0.2 < \eta < 0.7$ (``forward'', right panel) using EPS08 nPDFs with $k_T$ 
and HIJING plus $k_T$ with multiscattering. The two sets (EPS08 and HIJING) give
very similar data/theory ratios for both pseudorapidity intervals. There is
reasonable agreement with the E706 data~\cite{Apanasevich:2002wt} as is apparent 
from the quality of the data/theory ratios.

\begin{figure}[!h]
\begin{center} 
\includegraphics[width=8.5cm, height=8.5cm, angle=270]{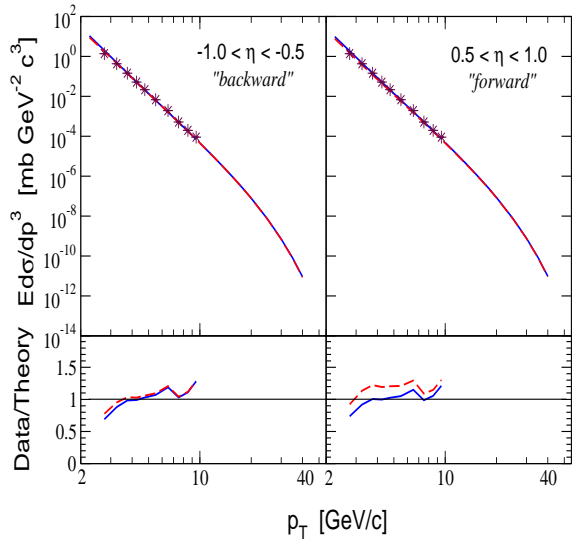}
\end{center}
\caption[...]{(Color Online) Spectra for $d+Au \rightarrow \pi{^0}+X$ at 
$-1.0 < \eta < -0.5$ (left panel) and $0.5 < \eta < 1.0$ (right panel). 
The solid line represents the
EPS08 with intrinsic $k_T$, while the dashed line is obtained from HIJING 
plus $k_T$ with the inclusion of multiscattering. Stars denote STAR data
averaged over $0.5 < |\eta| < 1.0$}
\label{fig:dAuspect2}
\end{figure}

The $dAu$ spectra are displayed in Fig.~\ref{fig:dAuspect2} for both 
EPS08 and HIJING at $-1.0 < \eta < -0.5$ (left panel) and $0.5 < \eta < 1.0$
(right panel). Note that the experimental data from the STAR 
collaboration~\cite{Abelev:2006pp} are not separated into negative and
positive pseudorapidities, but rather averaged over both 
intervals. Therefore, the data points in Fig.~\ref{fig:dAuspect2} only
serve to guide the eye; the relevance of the Figure is to highlight the 
difference between the EPS08 and HIJING results visible in the bottom panels.
It can be seen that the EPS08 results do not differ
much in the forward and backward directions, while HIJING gives significantly larger
data/theory ratios forward. Thus, EPS08 and HIJING 
differ appreciably at forward pseudorapidities, $0.5 < \eta < 1.0$.
This is not unexpected, because multiple scattering influences
the $d$-side (forward) more than the $Au$-side (backward).
To be able to draw stronger conclusions, separated forward and backward
data will be necessary, hopefully forthcoming from the high-statistics Run 8.

%
\subsection{Pseudorapidity asymmetry}
\label{etasmy}

\subsubsection{Asymmetry in $pBe$ collisions at $30.7$ GeV}
\label{asympBe}

\begin{figure}[!h]
\begin{center} 
\includegraphics[width=8.5cm, height=8.5cm, angle=270]{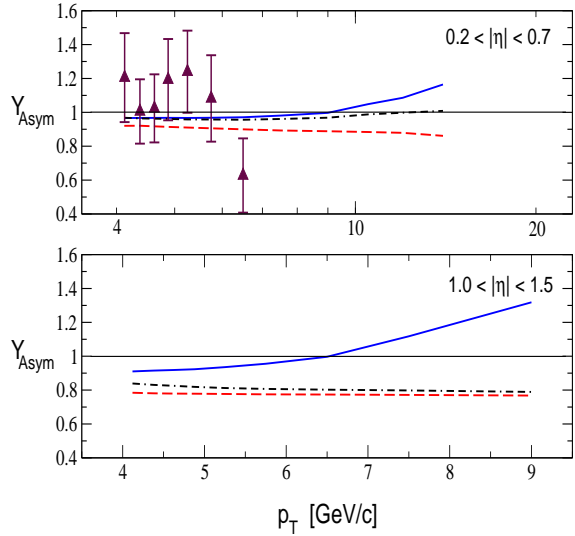}
\end{center}
\caption[...]{(Color Online) Pseudorapidity asymmetry,
$Y_{Asym}$ for $p+Be \rightarrow \pi{^0}+X$ at $0.2 < |\eta| < 0.7$ (top)
and  $1.0 < |\eta| < 1.5$ (bottom). The solid line represents the
EPS08 nPDFs, while the dashed line is obtained from HIJING with 
the inclusion of multiscattering. The dot-dashed line 
corresponds to HIJING without multiscattering, and filled triangles denote 
the E706 data~\cite{Apanasevich:2002wt}.}
\label{fig:asyfnal}
\end{figure}

The pseudorapidity asymmetry for the rapidity interval
$0.2 < |\eta| < 0.7$ is shown in the upper panel of Fig.~\ref{fig:asyfnal} 
compared with the E706 data~\cite{Apanasevich:2002wt}.
The solid line represents EPS08 with proton intrinsic $k_T$, the dot-dashed line 
displays HIJING with intrinsic $k_T$, and the dashed is HIJING plus 
intrinsic $k_T$ and multiple scattering. Both EPS08 with $k_T$ and HIJING 
without multiple scattering give very small asymmetries and the data are
also consistent with $Y_{Asym}=1$ for low $p_T$. The HIJING parameterization
with multiple scattering yields somewhat lower values at all transverse momenta.
This is in line with the effect of multiple scattering moderately increasing 
the yield on the $p$-side relative to that of the $Be$-side. In view of the rather 
large error bars, all three sets are in reasonable agreement with the data. 
The lower panel is our prediction for the interval $1.0 < |\eta| < 1.5$. 
Here, the calculated effects are larger, but have a similar structure to
what is seen at lower $\eta$. This trend is similar to what will be seen at 
other energies.

\subsubsection{Asymmetry in $dAu$ collisions at $200$ $A$GeV}
\label{asymdAu}

\begin{figure}[!h]
\begin{center} 
\includegraphics[width=8.5cm, height=8.5cm, angle=270]{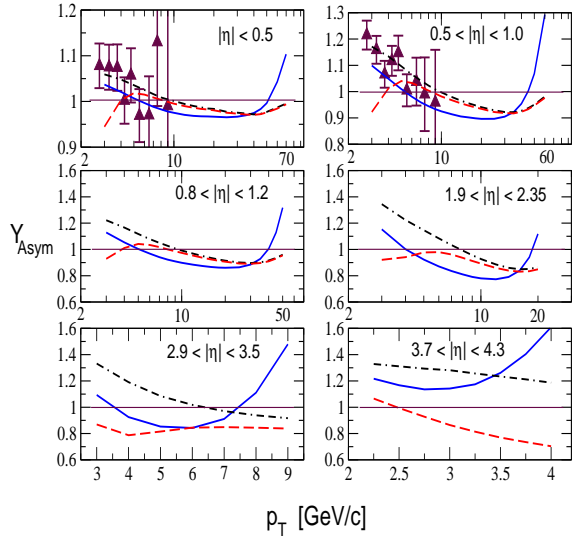}
\end{center}
\caption[...]{(Color Online) Pseudorapidity asymmetry,
$Y_{Asym}$ for $d+Au \rightarrow \pi{^0}+X$ at different 
pseudorapidity intervals. The solid line represents the
EPS08 nPDFs, while the dashed line is obtained using HIJING shadowing with 
the inclusion of multiscattering. The dot-dashed line 
corresponds to HIJING without multiscattering, and filled triangles denote 
the STAR data~\cite{Abelev:2006pp}.}
\label{fig:asyrhic}
\end{figure}

Figure~\ref{fig:asyrhic} shows the pseudorapidity asymmetry for $\pi^0$ 
production from $dAu$ collisions at RHIC, for different pseudorapidity intervals. 
The two uppermost panels are our results for the asymmetry at $|\eta| < 0.5$ 
and $0.5 < |\eta| < 1.0$ compared with the STAR data~\cite{Abelev:2006pp}. For
$p_T > 4.0$ GeV/c, the agreement with data is quite good for all three sets.
At lower $p_T$, multiple scattering increases the calculated yield mostly in the 
forward direction as discussed in Sec.~\ref{shad_multiscatt}, leading to 
asymmetries below unity. At very high $p_T$ we observe a divergence in 
the model predictions.
The lower four panels are our predictions for the asymmetry as pseudorapidity 
increases. The first three of these correspond to the BRAHMS
pseudorapidity intervals~\cite{Arsene:2004ux}. The general trend is that the
asymmetry becomes larger as $\eta$ increases. This mainly arises
from the strong shadowing in the larger nucleus at lower $x$ values. Also, 
since increasing $\eta$ leads to decreasing accessible $p_T$ 
due to phase space constraints, the effects of multiple scattering become 
more pronounced. In fact, for the largest $\eta$ considered, 
$3.7 < |\eta| < 4.3$, there is a marked difference between HIJING with 
and without multiple scattering for all $p_T$ considered. 

It is pertinent at this point to make some remarks about
pseudorapidity asymmetry in the BRAHMS pseudorapidity intervals.
The BRAHMS Collaboration~\cite{Arsene:2004ux} has observed a progressive 
suppression of both minimum bias nuclear modification $R_{dAu}$ and 
central-to-peripheral ratios, $R_{CP}$, with increasing
$\eta$. The present study is limited to minimum bias pseudorapidity
asymmetry, and both shadowing parameterizations (EPS08 and HIJING) adequately 
describe the existing experimental data at very forward rapidities. 
The EPS08 parameterization incorporates RHIC data at large rapidities, i.e. at
low $x$, and thus reproduces the data. Therefore, at least in the 
minimum bias case, shadowing seems sufficient for a good description of 
the suppression observed at low $x$. 
The situation is different for the geometry-dependent $R_{CP}$, 
where ($b$-independent) shadowing plus conjectured 
impact parameter dependencies~\cite{Adeluyi:2008qk,Vogt:2004hf} 
are clearly inadequate in describing the observed suppression.

\subsubsection{Asymmetry in $dPb$ collisions at $8.8$ $A$TeV}
\label{asymdPb}

\begin{figure}[!h]
\begin{center} 
\includegraphics[width=8.5cm, height=8.5cm, angle=270]{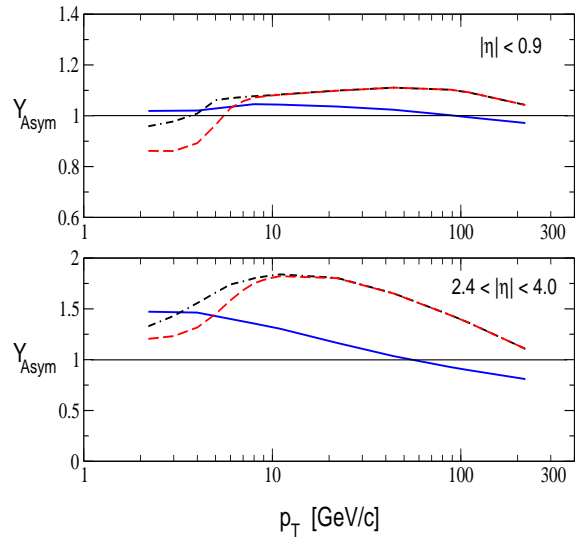}
\end{center}
\caption[...]{(Color Online) Predicted pseudorapidity asymmetry,
$Y_{Asym}$ for $d+Pb \rightarrow \pi{^0}+X$ at $\sqrt{s} = 8.8$ $A$TeV
for $|\eta| < 0.9$ and $2.4 < |\eta| < 4.0$. The solid line represents the
EPS08 nPDFs, while the dashed line is obtained from HIJING with 
the inclusion of multiscattering. The dot-dashed line 
corresponds to HIJING without multiscattering.}
\label{fig:asylhc}
\end{figure}

Let us now turn to our predictions for the pseudorapidity asymmetry in a potential future 
$dPb$ collision at LHC energy of $8.8$ $A$TeV. 
The calculated results are displayed in 
Fig.~\ref{fig:asylhc}, where the upper panel is for the interval $|\eta| < 0.9$ 
and the lower panel is for $2.4 < |\eta| < 4.0$. These 
intervals correspond to acceptance in the central detector and in the muon 
arm, respectively, of the ALICE experiment~\cite{Alessandro:2006yt}. All three 
sets predict minimal asymmetry of the order of a few percent for the interval 
$|\eta| < 0.9$. 
As we move to higher $\eta$, the predicted asymmetry becomes more significant. 
As can be seen in the lower panel of Fig.~\ref{fig:asylhc}, both EPS08 and 
HIJING predict substantial asymmetry up to $\sim 10$ GeV/c, and both variants
of HIJING asymmetries remaining significant up to $\sim 100$ GeV/c, in 
contrast to EPS08. 

At the present level, neither model variant gives agreement with all aspects of 
the data: in an earlier calculation we have found that shadowing 
parameterizations which do not need to be augmented by a multiple scattering 
prescription~\cite{Eskola:1998df,Shad_HKN,Frankfurt:2003zd} have difficulty 
describing central-to-peripheral ratios at forward rapidity~\cite{Adeluyi:2008qk}. 
We have checked that this also holds for the EPS08 nPDFs. On the other hand,
the HIJING parameterization with multiscattering yields pseudorapidity 
asymmetries below unity at low transverse momenta. This deficiency may be cured 
by allowing an $\eta$-dependent multiscattering~\cite{gergely:2008}. 
%
%
\section{Conclusion}
\label{concl}

Here we give a concise summary of the results of our
calculations. We have demonstrated the usefulness of asymmetric 
(light-on-heavy) nuclear collisions at relativistic energies. As illustrated
in Sec.~\ref{shad_multiscatt}, the physical differences between forward- and 
backward-going produced particles arise from the different ranges of $x$ 
sampled (different shadowing) and different amount of multiple scattering.
This leads to observable pseudorapidity asymmetries
at some collision energies and transverse momenta.
We have considered the effects of nuclear shadowing 
and multiple scattering on pseudorapidity asymmetry for three asymmetric 
systems: $pBe$, $dAu$, and $dPb$ in a wide energy range.

To calibrate and fine-tune our model we first examined spectra of
produced neutral pions. We found that there are two avenues in the model for 
the reasonable description of these data: (i) HIJING shadowing, intrinsic
$k_T$, plus multiscattering, or (ii) some other nPDFs (we use EPS08 here) 
and intrinsic $k_T$ (but no additional multiscattering). We then calculated 
pseudorapidity asymmetries with these prescriptions.
Overall, the calculated asymmetries are in reasonable agreement with available 
experimental data. Intrinsic transverse momentum in the nucleon is
seen to be important at low $p_T$. Multiple scattering 
increases the yield in the ``forward'' or positive pseudorapidity region,
thus leading to a tendency for asymmetries less than unity at low $p_T$
in the scheme explicitely relying on multiple scattering,
at variance with the data. An LHC measurement at a high pseudorapidity 
and high $p_T$ (where $k_T$ effects no longer make a difference) may be 
able to distinguish between strong shadowing (as in the HIJING prescription) 
and an nPDF with a relatively weaker gluon suppression (like e.g. EPS08).

A major constraint in assessing pseudorapidity asymmetries is the limited 
availability of data for direct comparison with theoretical calculations. 
More data in asymmetric light-on-heavy collisions separated with respect
to positive and negative pseudorapidities are needed
to judge calculated pseudorapidity asymmetries. At RHIC, it is expected
that the high-statistic $dAu$ Run 8 is going to provide such a large data set.

\section{Acknowledgments}
\label{ack}

This work was supported in part by Hungarian OTKA PD73596,
T047050, NK62044, and IN71374, by the U.S. Department of Energy under
grant U.S. DOE DE-FG02-86ER40251, and jointly by the U.S. and Hungary under
MTA-NSF-OTKA OISE-0435701.

%

\end{document}